\documentclass[final,3p,times,onecolumn]{elsarticle}

\usepackage{lineno,hyperref}
\usepackage{epstopdf}
\usepackage{comment}
\usepackage{graphicx,subfig}
\usepackage{amsmath,bm,amssymb,mathtools}
\usepackage{comment}
\usepackage{multicol,multirow}
\usepackage{algorithm,algpseudocode}
\usepackage[version=4]{mhchem}
\usepackage{siunitx}
\usepackage{longtable,tabularx}
\usepackage{hyperref}
\usepackage{color}
\biboptions{numbers,sort&compress}

\graphicspath{{figures/}}

\newtheorem{remark}{Remark}

\modulolinenumbers[5]

\date{}

\journal{  }

\begin{document}

\begin{frontmatter}

\title{Outlier-robust Kalman filters with mixture correntropy} %\tnoteref{funding}}

\tnotetext[funding]{This work was supported by the National Natural Science Foundation of China (NSFC) under Grant 61473227 and 11472222.}
\tnotetext[funding]{Corresponding author: Hongwei Wang (tianhangxinxiang@163.com).}

\author[add1,add2]{Hongwei Wang}
\author[add2]{Wei Zhang}
\author[add2]{Junyi Zuo}
\author[add2]{Heping Wang}

\address[add1]{National Key Laboratory of Science and Technology on Communications, University of Electronic Science and Technology of China, Chengdu, People's Republic of China, 611731}
\address[add2]{School of Aeronautics, Northwestern Polytechnical University, Xi'an, People's Republic of China, 710072}

\begin{abstract}
  We consider the robust filtering problem for a nonlinear state-space model with outliers in measurements. To improve the robustness of the traditional Kalman filtering algorithm, we propose in this work two robust filters based on mixture correntropy, especially the double-Gaussian mixture correntropy and Laplace-Gaussian mixture correntropy. We have formulated the robust filtering problem by adopting the mixture correntropy induced cost to replace the quadratic one in the conventional Kalman filter for measurement fitting errors. In addition, a tradeoff weight coefficient is introduced to make sure the proposed approaches can provide reasonable state estimates in scenarios where measurement fitting errors are small. The formulated robust filtering problems are iteratively solved by utilizing the cubature Kalman filtering framework with a reweighted measurement covariance. Numerical results show that the proposed methods can achieve a performance improvement over existing robust solutions.
\end{abstract}


\end{frontmatter}

%\linenumbers

\section{Introduction}

State estimation for stochastic discrete-time dynamic systems is one of the vital issues in control engineering, and it has broad applications in various areas, such as target tracking, sparse signal processing, fault detection and diagnosis, pose estimation, and many others~\cite{grewal2010applications,auger2013industrial,lu2015adaptive,wang2016variational,lou2017desensitized,hashim2018nonlinear,hashim2019nonlinear}. The state estimates of a linear system with Gaussian noises is provided by the celebrated Kalman filter~\cite{kalman1960new}. For nonlinear systems with a Gaussian assumption (i.e., both the process and measurement noises are Gaussian), several Kalman-liked Gaussian approximation filters (GKF) were investigated, e.g., the unscented Kalman filter~\cite{julier2004unscented}, cubature Kalman filter~\cite{arasaratnam2009cubature,wang2017generalized}, to name a few. These solutions have shown good performance when the Gaussian assumption meets in systems. In some applications, however, the Gaussian assumption of the measurement noise may fail since outliers may contaminate measurements due to unreliable sensors. Outliers lead to the measurement noise having a heavy tail and becoming non-Gaussian, resulting in substantial degradation of the existing GKFs.

The sequential Monte-Carlo sampling/particle filter (PF)~\cite{arulampalam2002tutorial} and the Gaussian sum filter (GSM) are two general strategies to deal with non-Gaussian noises caused by measurement outliers. In the PF, a massive number of particles are involved to approximate the posterior probability density function to obtain reasonable estimation results. In the GSM, the state estimates are obtained by combining the results from several parallelly implemented filters via an interacting procedure. Therefore, both the PF and GSM suffer from a great computational burden, which prevents them from being widely used in applications. In addition, a computationally economical approach, i.e., integrating the robust cost from $M$-estimation (e.g., Huber's cost) into the GKF framework~\cite{karlgaard2007huber,karlgaard2014nonlinear,chang2017unified}, has also been studied. This type of robust filters was developed by interpreting the GKF filtering problem as linear or nonlinear regression. Other approaches for robust filtering such as the heavy-tailed distribution based solution~\cite{wang2017laplace} and the $H_{\infty}$ filter~\cite{luan2010h} were also reported in the literature.

Recently, a novel local similarity measure called correntropy from information-theoretic learning is introduced to deal with heavy-tailed non-Gaussian noises~\cite{liu2007correntropy,ma2015maximum,luo2018towards}, and its associated maximum correntropy criterion (MCC) has been employed to design a robust filtering algorithm. In~\cite{cinar2012hidden,izanloo2016kalman} the MCC was first employed to improve the robustness of the KF for linear systems. Those methods employed the gradient descent approach and ignored the covariance propagation procedure, which may cause a potential loss of information. To handle this issue, a robust Kalman filter called MCKF~\cite{chen2017maximum} was developed via recasting the Kalman filtering problem as a linear regression one. Afterward, several variants of the MCKF were developed for nonlinear systems~\cite{wang2017maximum,kulikova2017square,wang2018maximum}. Although the feasibility of the MCC based robust filters for dealing with non-Gaussian noises has been demonstrated, the default kernel in the MCC, e.g., the Gaussian kernel, may not sufficient to deal with more complex data in many practical problems~\cite{chen2018mixture,wang2019robust}. Besides, there is still no guideline for the selection of the kernel parameter which has a significant influence of the MCC associated robust filter.

There is a need, therefore, for designing some kernel parameter-insensitive algorithms to deal with measurement outliers. To address this challenge, we propose two robust Kalman filters based on the mixture correntropy. We formulate the robust filtering problem by utilizing a mixture correntropy induced loss to replace the quadratic one in the GKF for measurement fitting errors. In addition, a weighting coefficient is included to seek a tradeoff between the model and measurement fitting errors. The resulting robust filtering problem are iteratively solved within the GKF framework with a reweighted measurement covariance. The simulation results have shown the superior performance of the proposed algorithms, and also provide a heuristic rule to design kernel parameters.

The remaining of the paper is organized as follows. In Section~\ref{s2}, we give a brief introduction of mixture correntropy. In section~\ref{s3}, we formulate the mixture correntropy based robust Kalman filtering problems and derive the related algorithms. In Section~\ref{sim} we present a simulation example to verify the performance of the proposed algorithm. Finally, Section~\ref{s5} concludes our work.

\emph{Notation}: In this paper, boldface lower and upper-case letters represent column vectors and matrices, respectively. Scalars are denoted by normal font letters. $A^T$ means the transpose of the matrix $A$. $\mathcal{N}(\cdot,\cdot)$ means a Gaussian distribution. The estimates of state $\bm x_t$ given the measurements up to $t=m$ is denoted by $\hat{\bm x}_{t|m}$.

\section{Brief review of the mixture correntropy}
\label{s2}

Correntropy is a newly developed similarity measure that originated from information-theoretic learning. Given two random variables $X$ and $Y$, correntropy is defined by~\cite{liu2007correntropy}
\begin{align}
V(X,Y) = E[\kappa(X,Y)]=\int\int{\kappa(X,Y)p(x,y)dxdy}
\end{align}
where $\kappa(\cdot,\cdot)$ is a kernel function which satisfies Mercer's theorem, $E(\cdot)$ denotes the expectation operation and $p(x,y)$ is the joint probability density function of $X$ and $Y$. Readers can refer~\cite{liu2007correntropy} for the properties of correntropy and its associated maximum correntropy criterion. The advantages of correntropy in dealing with measurement outliers in the Kalman filtering framework were illustrated in several literatures, e.g.,~\cite{cinar2012hidden,izanloo2016kalman,chen2017maximum,wang2017maximum,kulikova2017square,wang2018maximum}. However, correntropy with a single kernel may suffer performance degradation when dealing with more complex data. In addition, correntropy is sensitive to the parameter of the kernel function, which may limit its performance. To address those issues and improve the flexibility, a mixture correntropy is introduced~\cite{chen2018mixture}, i.e.,
\begin{align}
M(X,Y)=E\left(\alpha\kappa_1(X,Y)+(1-\alpha)\kappa_2(X,Y)\right)
\label{mix_co}
\end{align}
where $0<\alpha<1$ is the mixture coefficient, $\kappa_1(\cdot)$ and $\kappa_2(\cdot)$ are two different Mercer kernel functions. Calculating the exact value of the mixture correntropy is in general intractable due to the lack of knowledge of $p(x,y)$. In practice, only some finite data samples $\{x_i,y_i\}_{i=1}^N$ are available, hence the value of mixture correntropy can be empirically approximated by
\begin{align}
M(X,Y)=\frac{1}{N}\sum_{i=1}^N\left(\alpha \kappa_1(e_i)+(1-\alpha)\kappa_2(e_i))\right)
\end{align}
where $e_i=x_i-y_i$. It is clear that the mixture correntropy will reduce to the original correntropy when $\alpha=0$ or $\alpha=1$.

\begin{remark}
For simplicity, we mainly focus on the mixture correntropy with two different kernels in this work. The mixture correntropy, however, has a generalized form other than the definition in~\eqref{mix_co}, i.e., $M(X,Y)=E\left(\sum_i\alpha_i\kappa_i(X,Y)\right)$ with $i\ge 2$, $\alpha_i>0$ and $\sum_i\alpha_i=1$.
\end{remark}

Generally, the difference between $\kappa_1(\cdot)$ and $\kappa_2(\cdot)$ in the mixture correntropy can be reached via two approaches. In the first one, $\kappa_1(\cdot)$ and $\kappa_2(\cdot)$ may come from the same kernel family but with distinct kernel parameters, resulting in a homogenous mixture correntropy, e.g., the double-Gaussian kernel mixture correntropy (DG-MC)~\cite{chen2018mixture} where
\begin{align*}
\kappa_1(e_i)  = \exp(-\frac{e_i^2}{2\sigma_1^2})\ ,\
\kappa_2(e_i)  = \exp(-\frac{e_i^2}{2\sigma_2^2})
\end{align*}
In the other, $\kappa_1(\cdot)$ and $\kappa_2(\cdot)$ may be the different types of kernel functions, leading to a heterogenous mixture correntropy, e.g., the Laplace-Gaussian kernel mixture correntropy (LG-MC) proposed in~\cite{wang2019robust} in which
\begin{align*}
\kappa_1(e_i)  = \exp(-\frac{e_i^2}{2\sigma_1^2})\ ,\
\kappa_2(e_i)  = \exp(-\frac{|e_i|}{\sigma_2})
\end{align*}

It is apparent that both the DG-MC and LG-MC meet their maximum when $X = Y$ (i.e., two random variables are exactly the same). Therefore, we here define the mixture correntropy loss in~\eqref{mcl} for facilitating the formulation of the optimization problem
\begin{align}
L(X,Y)=1-M(X,Y)
\label{mcl}
\end{align}
In the next section, we devote to utilizing the DG-MC loss (DG-MCL) and LG-MC loss (LG-MCL) to design robust filters.

\section{Derivation of the proposed robust Kalman filter}
\label{s3}

Consider the stochastic dynamic process described by a state-space model
\begin{align}
\bm x_t &= f(\bm x_{t-1})+\bm w_{t-1}\label{process}\\
\bm y_t &= h(\bm x_t) + \bm v_t\label{measurement}
\end{align}
where $\bm y_t \in \mathcal{R}^m$ is a measurement related to the state of interest $\bm x_t\in \mathcal{R}^n$; $f(\cdot)$ and $h(\cdot)$ are some known mappings to model the state transition and measurement procedure respectively; $\bm w_{t-1}\sim\mathcal{N}(0,\bm Q_{t-1})$ is the process noise and $\bm v_t$ is the measurement noise. In canonical Kalman filtering, $\bm v_t$ is assumed to be Gaussian, i.e., $\bm v_t\sim\mathcal{N}(0,\bm R_{t})$. Under such a Gaussian assumption, the Kalman filtering problem can be formulated as the following minimization problem
\begin{align}
\hat{\bm x}_{t|t} & = \arg\min_{\bm x_t} \ -\log p(\bm x_t|\bm y_{1:t})\notag\\
& =  \arg\min_{\bm x_t}\left( -\log p (\bm x_t|\bm y_{1:t-1}) - \log p(\bm y_t|\bm x_t)\right)
\label{opt111}
\end{align}
where $p(\bm y_t|\bm x_t)$ is the likelihood function given by $\mathcal{N}(h(\bm x_t),\bm R_t)$, and $ p (\bm x_t|\bm y_{1:t-1})$ is the predictive density which can be approximated by $\mathcal{N}(\hat{\bm x}_{t|t-1},\bm P_{t|t-1})$ in the Gaussian approximation filtering framework. Substituting both the predictive density and likelihood distribution into~\eqref{opt111}, and discarding the terms that do not depend on $\bm x_t$, we can rewrite\eqref{opt111} as
\begin{align}
\hat{\bm x}_{t|t} &= \arg\min_{\bm x_t}\left(\frac{1}{2}\|\bm x_t-\hat{\bm x}_{t|t-1}\|_{\bm P_{t|t-1}^{-1}}^2+\frac{1}{2}\|\bm y_t-h(\bm x_t)\|_{\bm R_t^{-1}}^2\right)\notag\\
&=\arg\min_{\bm x_t}\left(\frac{1}{2}\|\bm x_t-\hat{\bm x}_{t|t-1}\|_{\bm P_{t|t-1}^{-1}}^2+\frac{1}{2}\sum_{i=1}^m e_{t,i}^2\right)
\label{trad_filter}
\end{align}
where $e_{t,i}$ is the $i$-th component of $\bm e_t = \bm R_t^{-1/2}(\bm y_t-h(\bm x_t))$. The optimization problem in~\eqref{trad_filter} can be solved by several GKFs, e.g., CKF. In the following, we present our robust Kalman filters in conjunction with the CKF which is briefly introduced in~\ref{appe}. It is straightforward to extend the proposed filters with other GKFs.

From~\eqref{trad_filter} we note that the traditional Kalman filter based on the Gaussian assumption has a quadratic loss for the measurement fitting error. It is clear that the quadratic loss is sensitive to outliers, which is the main reason that causes the performance degradation of the KF in scenarios where measurement outliers encountered. In order to improve the robustness of the filtering algorithm against outliers, the DG-MCL and LG-MCL are utilized separately to replace the quadratic loss for the measurement fitting error to design the mixture correntropy based robust filters.

\subsection{DG-MCL based robust Kalman filter}

In this section, we first derive a robust Kalman filter based on the DG-MCL. Adopting the DG-MCL to the measurement fitting error leads to the following robust filtering problem
\begin{align}
\hat{\bm x}_{t|t} = \arg\min_{\bm x_t}\left\{\frac{1}{2}\|\bm x_t-\hat{\bm x}_{t|t-1}\|_{\bm P_{t|t-1}^{-1}}^2+\lambda\left[1-\frac{1}{m}\sum_{i=1}^m\left(\alpha\exp(-\frac{e_i^2}{2\sigma_1^2})+(1-\alpha)\exp(-\frac{e_i^2}{2\sigma_2^2})\right)\right]\right\}
\label{op_dg}
\end{align}
where $\lambda$ is a weighting coefficient to make the balance between the model fitting error and measurement fitting error. $\lambda$ should be carefully chosen to obtain a reasonable estimation result. Specifically, we expect that the performance of the DG-MCL is similar to that of the quadratic loss when the measurement fitting error is small. It is noticed that for a small real vale $\delta$, we have
\begin{align*}
e^{\delta}\approx 1 + \delta
\end{align*}
Therefore, for a small measurement fitting error, the DG-MCL can be approximated as
\begin{align*}
{L}_{DG-MCL}\approx \frac{\alpha\sigma_2^2+(1-\alpha)\sigma_1^2}{m\sigma_1^2\sigma_2^2}\frac{1}{2}\sum_{i=1}^m {e_{t,i}^2}
\end{align*}
In order to maintain the similarity of the quadratic loss and DG-MCL when the measurement fitting error is small, $\lambda$ should be determined as
\begin{align}
\lambda=\frac{m\sigma_1^2\sigma_2^2}{\alpha\sigma_2^2+(1-\alpha)\sigma_1^2}
\end{align}

Differencing the cost function in~\eqref{op_dg} with regards to $\bm x_t$, we have
\begin{align}
\bm P_{t|t-1}^{-1}(\bm x_t-\hat{\bm x}_{t|t-1})-\frac{\lambda}{m}\sum_{i=1}^m\left(\alpha\frac{\partial \kappa_1(e_i)}{\partial e_i}\frac{\partial e_i}{\partial \bm x_t}+
(1-\alpha)\frac{\partial \kappa_2(e_i)}{\partial e_i}\frac{\partial e_i}{\partial \bm x_t}\right)=0
\label{op_dg2}
\end{align}
For the Gaussian kernel, we know that
\begin{align}
\frac{\partial \kappa_1(e_i)}{\partial e_i}=-\frac{e_i}{\sigma_1^2}\kappa_1(e_i),\quad
\frac{\partial \kappa_2(e_i)}{\partial e_i}=-\frac{e_i}{\sigma_2^2}\kappa_2(e_i)
\label{d1}
\end{align}
Substituting~\eqref{d1} into~\eqref{op_dg2} results in
\begin{align}
\bm P_{t|t-1}^{-1}(\bm x_t-\hat{\bm x}_{t|t-1})+\frac{\lambda}{m}\sum_{i=1}^{m}\left(\frac{\alpha \kappa_1(e_i)}{\sigma_1^2}+\frac{(1-\alpha)\kappa_2(e_i)}{\sigma_2^2}\right)\frac{e_i\partial e_i}{\partial \bm x_t} = 0
\label{op_dg23}
\end{align}
Define a diagonal matrix $\bm \Lambda_t$ with its $i$-th element given by
\begin{align}
\Lambda_{t,ii}=\frac{\lambda}{m}\left(\frac{\alpha \kappa_1(e_i)}{\sigma_1^2}+\frac{(1-\alpha)\kappa_2(e_i)}{\sigma_2^2}\right)
\label{lambda}
\end{align}
With $\bm \Lambda_t$, one can rewrite~\eqref{op_dg23} into the matrix format as
\begin{align}
\bm P_{t|t-1}^{-1}(\bm x_t-\hat{\bm x}_{t|t-1})+\frac{\partial \bm e_t}{\partial \bm x_t}\bm \Lambda_t\bm e_t=0
\label{tti}
\end{align}
Equation~\eqref{tti} is essentially the derivative of the cost function of the following optimization problem
\begin{align}
\hat{\bm x}_{t|t}=\arg\min_{\bm x_t}\left(\frac{1}{2}\|\bm x_t-\hat{\bm x}_{t|t-1}\|_{\bm P_{t|t-1}^{-1}}^2+\frac{1}{2}\|\bm y_t-h(\bm x_t)\|_{\bar{\bm R}_t^{-1}}^2\right)
\label{op_dg3}
\end{align}
where
\begin{align}
\bar{\bm R}_t=\bm R_t^{T/2}\bm\Lambda_t^{-1}\bm R_t^{1/2}
\label{rt}
\end{align}

Despite simple structure, directly solving~\eqref{op_dg3} is intractable due to the fact that $\bar{\bm R}_t$ depends on the state $\bm x_t$ via $\bm \Lambda_t$. To address this, we adopt an {\bf{alternate iterative algorithm}}. Specifically, for the given estimate $\hat{\bm x}_{t|t}^{k}$ after the $k$-th iteration, we construct $\bm\Lambda^{k}_t$ via~\eqref{lambda}, and then $\bar{\bm R}_t^{k}$ via~\eqref{rt}. In the next iteration, we solve the optimization problem~\eqref{op_dg3} with $\bar{\bm R}_t^{k}$ to obtain $\hat{\bm x}_{t|t}^{k+1}$. It is noted that~\eqref{op_dg3} has a similar structure as the one under the Gaussian assumption illustrated in~\eqref{trad_filter}, which enables us to solve~\eqref{op_dg3} by applying the existing Gaussian approximation filtering solutions, e.g., the CKF in~\ref{appe}. This iteration loop continues until the algorithm converges, e.g., for a small tolerance $\epsilon$,
\begin{align}
\|\hat{\bm x}_{t|t}^{k+1}-\hat{\bm x}_{t|t}^{k}\|<\epsilon
\end{align}

At the beginning of the iteration procedure, we initialize $\bm\Lambda_t$ as an identity matrix, meaning that in the first loop the conventional CKF is implemented. The proposed robust filter is summarized in Algorithm~\ref{alg2}.

\begin{algorithm}[!h]
  \caption{DG-MCL based robust CKF (DG-MCL-CKF)} \label{alg2}
  \begin{algorithmic}[0]
    \State \textbf{Input:} $\bm {y}_{1:T}$, $\hat{\bm x}_{0|0}$, $\bm P_{0|0}$, $\bm Q_{1:T}$, $\bm R_{1:T}$,$\sigma_1$,$\sigma_2$.
    \State \textbf{Output:} $\bm \hat{\bm x}_{t|t}$ and $\bm P_{t|t}$ for $t=1:T$.
    \For {$t=1:T$}
    \State Update $\{\hat{\bm x}_{t|t-1},{\bm P}_{t|t-1}\}$ via \{\eqref{pu_1},\eqref{pu_2}\};
    \State Initialize $k=0$, $\bm \Lambda_{t}=\bm I_m$;
    \Repeat {\quad $k = 1,\cdots,$}
    \State Update $\bar{\bm R}_t$ via~\eqref{rt} using $\bm \Lambda_{t}$;
    \State Update $\hat{\bm x}_{t|t}^k$ and $\bm P_{t|t}^k$ via~\eqref{x_upt} and~\eqref{p_upt} respectively;
    \State Calculate $\bm e_t = \bm R_t^{-1/2}(\bm y_t-h(\hat{\bm x}_{t|t}^k))$, and update $\bm \Lambda_{t}$ via~\eqref{lambda};
    \State Calculate $\epsilon=\|\hat{\bm x}_{t|t}^{k+1}-\hat{\bm x}_{t|t}^{k}\|$;
    \Until {$\epsilon< 10^{-6}$}
    \State $\hat{\bm x}_{t|t} = \hat{\bm x}_{t|t}^k$, ${\bm P}_{t|t}={\bm P}_{t|t}^k$.
  \EndFor
  %--------------- for loop -----------------------
  \end{algorithmic}
\end{algorithm}

\subsection{LG-MCL based robust Kalman filter}

Similarly, the LG-MCL based robust filtering problem can be formulated as
\begin{align}
\hat{\bm x}_{t|t} = \arg\min_{\bm x_t}\Bigg\{\frac{1}{2}\|\bm x_t&-\hat{\bm x}_{t|t-1}\|_{\bm P_{t|t-1}^{-1}}^2+\left[1-\frac{1}{m}\sum_{i=1}^m\lambda_i\left(\alpha\exp(-\frac{e_i^2}{2\sigma_1^2})+(1-\alpha)\exp(-\frac{|e_i|}{\sigma_2})\right)\right]\Bigg\}
\label{op_lg}
\end{align}
where $\lambda_i$ is the weighting parameter. The reason why we assign a weighting parameter to each component of the measurement fitting is that the kernel functions in the LG-MC are heterogenous. Likewise, for a small fitting error, the LG-MCL can be approximated by
\begin{align}
{L}_{DG-MCL}\approx \frac{1}{2}\sum_{i=1}^m \eta_i e_{i,t}^2
\end{align}
where $\eta_i$ is given by
\begin{align*}
\eta_i = \frac{1}{m}(\frac{\alpha}{\sigma_1^2}+\frac{2(1-\alpha)}{\sigma_2|e_i|})
\end{align*}
Therefore, $\lambda_i$ should be chosen as in~\eqref{lbdi} to make sure that the LG-MCL has a similar performance of the quadratic loss when dealing with the small measurement fitting error
\begin{align}
\lambda_i = \eta_t^{-1} = m(\frac{\alpha}{\sigma_1^2}+\frac{2(1-\alpha)}{\sigma_2|e_i|})^{-1}
\label{lbdi}
\end{align}

Akin to the derivation of the DG-MCL based robust Kalman filter, the reformulated optimization problem for the LG-MCL based robust filtering problem is
\begin{align}
\hat{\bm x}_{t|t}=\arg\min_{\bm x_t}\left(\frac{1}{2}\|\bm x_t-\hat{\bm x}_{t|t-1}\|_{\bm P_{t|t-1}^{-1}}^2+\frac{1}{2}\|\bm y_t-h(\bm x_t)\|_{\bar{\bm R}_t^{-1}}^2\right)
\label{op_lg3}
\end{align}
where
\begin{align}
\bar{\bm R}_t&=\bm R_t^{T/2}\bm\Lambda_t^{-1}\bm R_t^{1/2}\label{lgrt}\\
\bm\Lambda_t&=\text{diag}(\Lambda_{t,11},\cdots,\Lambda_{t,mm})\label{lbd}\\
\Lambda_{t,ii}&= \frac{\lambda_i}{m}\left(\frac{\alpha}{\sigma_1^2}\exp(-\frac{e_i^2}{2\sigma_1^2})+\frac{2(1-\alpha)}{|e_i|\sigma_2}\exp(-\frac{|e_i|}{\sigma_2})\right)\label{lbd2}
\end{align}

Here we apply the similar iterative procedure to solve~\eqref{op_lg3}, and the details of the resulting robust filter is presented in Algorithm~\ref{alg1}.

\begin{algorithm}[!h]
  \caption{LG-MCL based robust CKF (LG-MCL-CKF)} \label{alg1}
  \begin{algorithmic}[0]
    \State \textbf{Input:} $\bm {y}_{1:T}$, $\hat{\bm x}_{0|0}$, $\bm P_{0|0}$, $\bm Q_{1:T}$, $\bm R_{1:T}$,$\sigma_1$,$\sigma_2$.
    \State \textbf{Output:} $\bm \hat{\bm x}_{t|t}$ and $\bm P_{t|t}$ for $t=1:T$.
    \For {$t=1:T$}
    \State Update $\{\hat{\bm x}_{t|t-1},{\bm P}_{t|t-1}\}$ via \{\eqref{pu_1},\eqref{pu_2}\};
    \State Initialize $k=0$, $\bm \Lambda_{t}=\bm I_m$;
    \Repeat {\quad $k = 1,\cdots,$}
    \State Update $\bar{\bm R}_t$ via~\eqref{lgrt} using $\bm \Lambda_{t}$;
    \State Update $\hat{\bm x}_{t|t}^k$ and $\bm P_{t|t}^k$ via~\eqref{x_upt} and~\eqref{p_upt} respectively;
    \State Calculate $\bm e_t = \bm R_t^{-1/2}(\bm y_t-h(\hat{\bm x}_{t|t}^k))$, and update $\bm \Lambda_{t}$ via~\eqref{lbd};
    \State Calculate $\epsilon=\|\hat{\bm x}_{t|t}^{k+1}-\hat{\bm x}_{t|t}^{k}\|$;
    \Until {$\epsilon< 10^{-6}$}
    \State $\hat{\bm x}_{t|t} = \hat{\bm x}_{t|t}^k$, ${\bm P}_{t|t}={\bm P}_{t|t}^k$.
  \EndFor
  %--------------- for loop -----------------------
  \end{algorithmic}
\end{algorithm}

\section{Simulations and results}
\label{sim}

In this section, we analyze the proposed algorithms by investigating two numerical simulations, e.g., estimating the state of a Van der Pol oscillator (VPO) and the state-of-charge (SoC) of a battery. For comparison, we also consider the conventional CKF and some existing robust filters, including the maximum correntropy derivative-free robust CKF (MCC-CKF)~\cite{wang2018maximum}, linear regression and maximum correntropy based CKF (RMCC-CKF)~\cite{wang2017maximum} and Huber's cost function based CKF (Huber-CKF)~\cite{chang2017unified}. We use two different setups in the MCC-CKF, i.e., the MCC-CKF1 with $\{\sigma=100,\ \eta=4\}$ and MCC-CKF2 with $\{\sigma=100,\ \eta=5\}$ (setting $\sigma=100$ is to deal with the Gaussian process noise). We set the kernel parameter in the RMCC-CKF to $5$ and the threshold parameter in the Huber-CKF to $1.345$. In the DG-MCL-CKF and LG-MCL-CKF, kernel parameters are determined as $\sigma_1=4$ and $\sigma_2=5$, and the mixture coefficient $\alpha$ is set to $0.5$.

\begin{remark}
The kernel parameters in the mixture correntropy will influence the performance of the MCL. In this work, we select these user-defined parameters by trial and error. Further studies, however, are needed to explore the detailed parameter-selection strategy, which would be beyond the scope of this work.
\end{remark}

\subsection{VPO model}

The standard VPO model is given by
\begin{align*}
\left\{\begin{array}{l}
\dot{x}_1=x_2\\
\dot{x}_2=\mu(1-x_1^2)x_2-x_1
\end{array}
\right.
\end{align*}
where $\mu$ is a coefficient to control the nonlinearity of the VPO. Using a sampling interval $\delta$ to discretize the VPO results
\begin{align}
\bm x_{t} = \bm x_{t-1}+\left(\begin{array}{l}
\int_{\delta}x_2dt\\
\int_{\delta}(\mu(1-x_1^2)x_2-x_1)dt
\end{array}
\right) + \bm w_{t-1}
\label{vpo_pro}
\end{align}
where $\bm x_t=[x_{1,t},x_{2,t}]^T$ is the state of interest and $\bm w_t$ is the process noise which is assumed to be Gaussian, i.e., $\bm w_t\sim\mathcal{N}(0,\bm Q_{t-1})$. We utilize the fourth-order Runge-Kutta scheme to numerically calculate the integral terms in~\eqref{vpo_pro} which in general have no analytical solutions. Furthermore we assume that the noisy measurements are gathered via
\begin{align*}
y_t = (x_{1,t}-1)^2+1+v_t
\end{align*}
The measurement noise is modeled as the following Gaussian-mixture model to simulate the heavy-tailed property caused by outliers
\begin{align*}
v_t\sim(1-\phi)\mathcal{N}(0,R_t)+\phi\mathcal{N}(0,\varphi R_t)
\end{align*}
in which $\phi$ is the contaminating ratio, $\varphi$ is the outlier strength factor and $R_t$ is the covariance of the nominal measurement noise.

In the simulation, we set $\mu=1$, and total samples $T=120$ with the sampling interval $\delta=0.1$s are involved. The true value of the initial state is $\bm x_0=[0,-0.5]^T$ and the estimated initial state is generated by a Gaussian distribution $\mathcal{N}({\bm x}_{0},0.01\bm I_2)$. The covariance of the process noise and the nominal measurement noise are, respectively, given by $\bm Q_{t-1}=0.005\bm I_2$ and $R_t=1$. $L=1000$ Monte Carlo runs are implemented to obtain the simulation results. The time-averaged root mean square (TRMSE) is employed as a metric, which is defined as
\begin{align*}
\text{TRMSE}_k=\frac{1}{T}\sum_{t=1}^T\sqrt{\frac{1}{L}\sum_{i=1}^L(x_{k,t}^i-\hat{x}_{k,t}^i)^2},\quad k=1,2
\end{align*}

\begin{figure*}[th]
\centering
\subfloat[TRMSE of $x_1$]{\includegraphics[width=0.49\columnwidth]{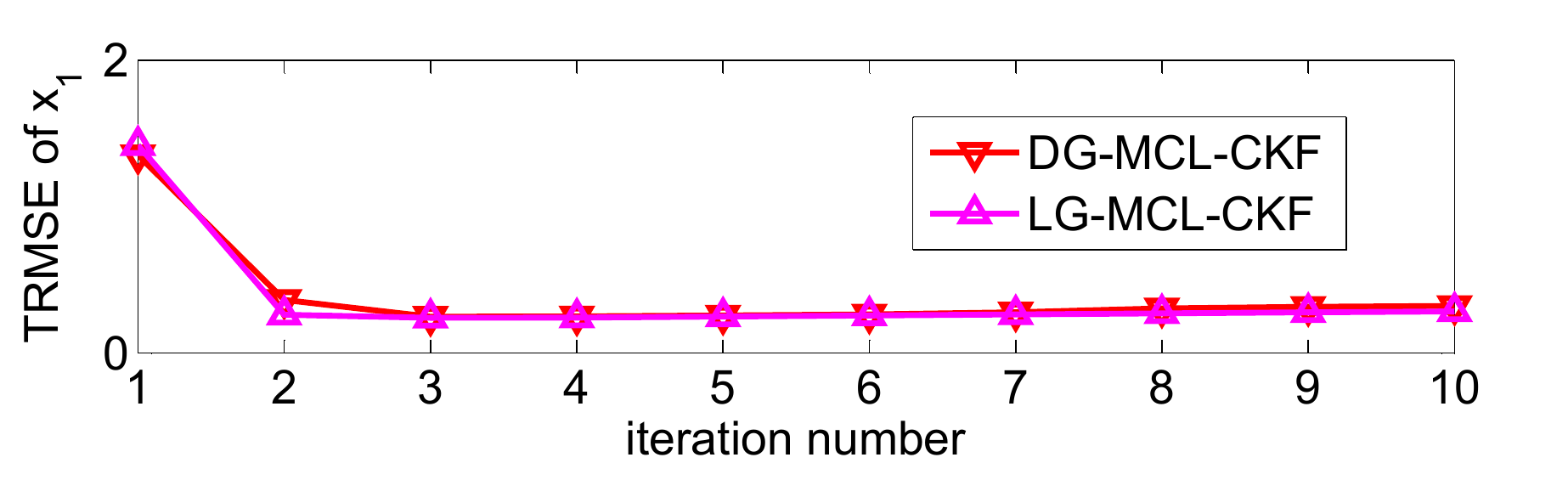}}
\subfloat[TRMSE of $x_2$]{\includegraphics[width=0.49\columnwidth]{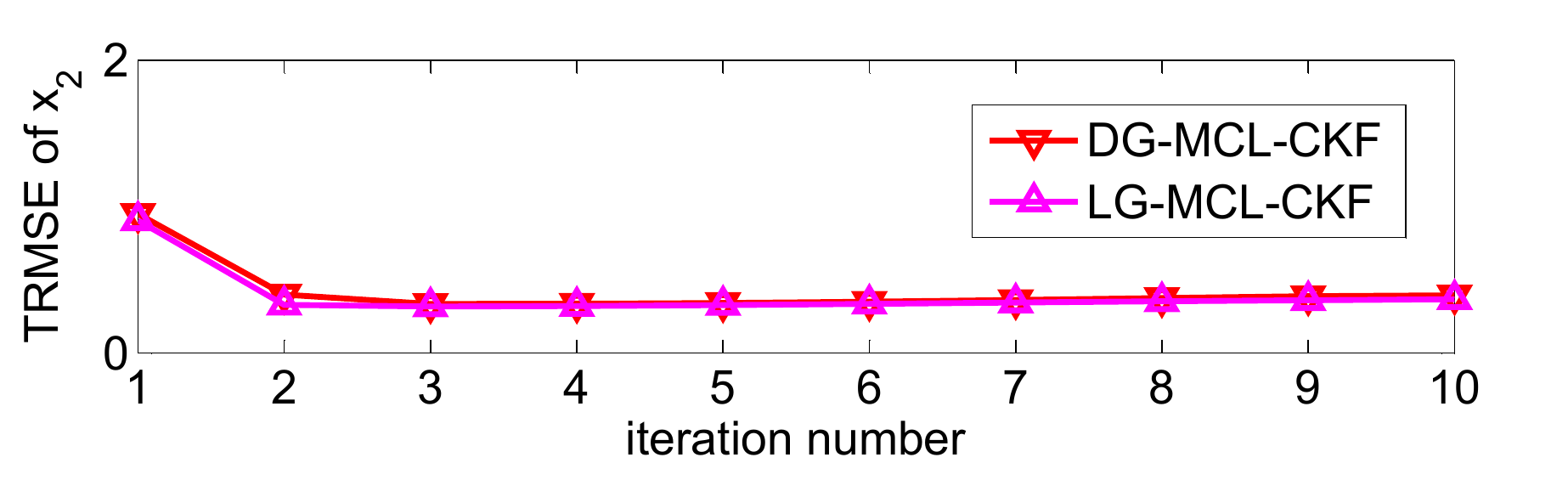}}
\caption{TRMSEs versus iteration number when $\varphi=200$ and $\phi=0.1$.}
\label{fff1}
\end{figure*}

First, we have studied the performance of the proposed methods versus the iteration number. Fig.~\ref{fff1} shows the TRMSEs of $x_1$ and $x_2$ when the iteration number of our algorithms varies from $1$ to $10$. It is apparent that the proposed approaches converge after $2$ or $3$ iterations. In the following simulations, we set $3$ as a default value of the iteration number for the proposed algorithms.

\begin{figure*}[th]
\centering
\subfloat[TRMSE of $x_1$]{\includegraphics[width=0.49\columnwidth]{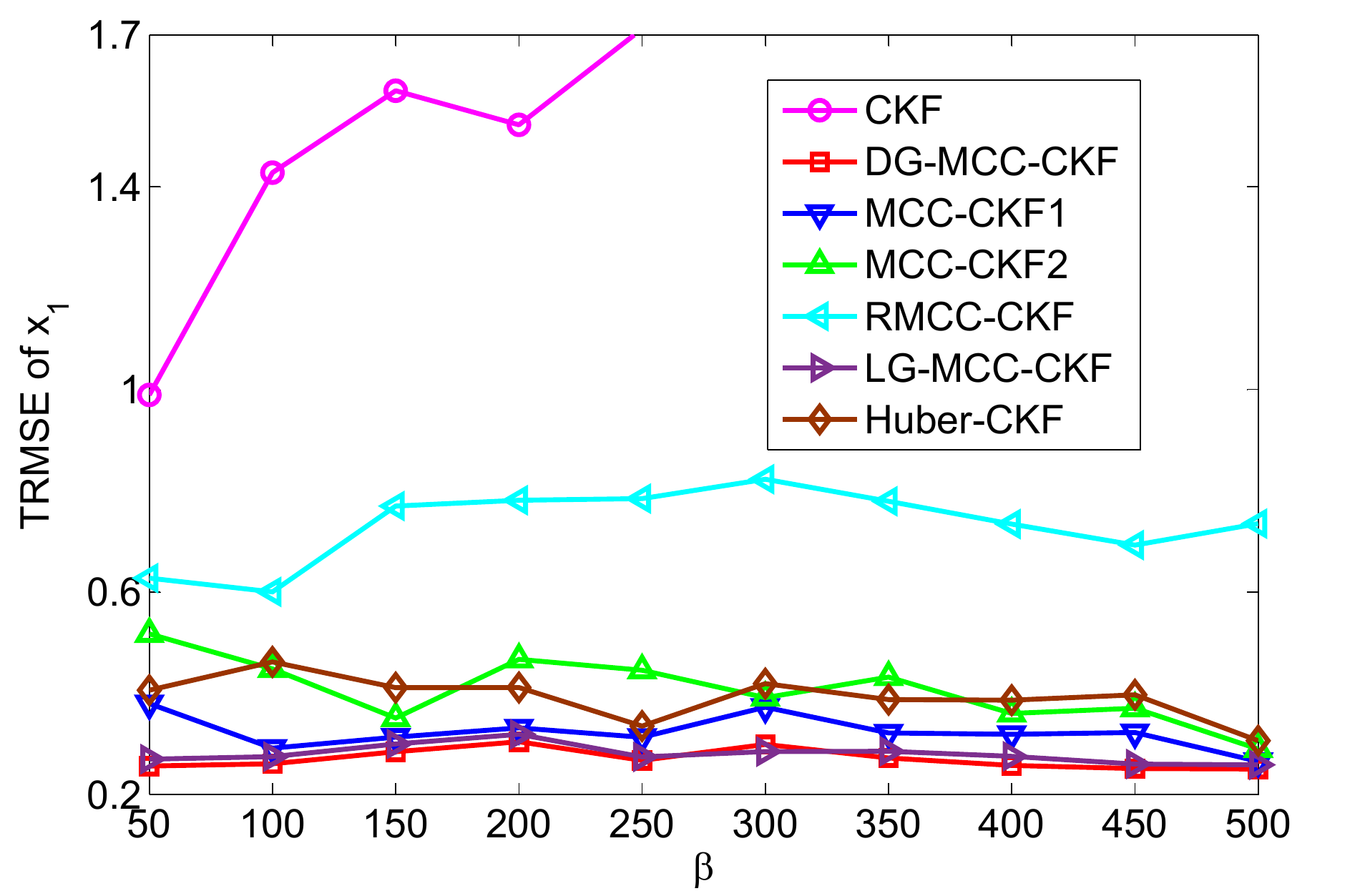}}
\subfloat[TRMSE of $x_2$]{\includegraphics[width=0.49\columnwidth]{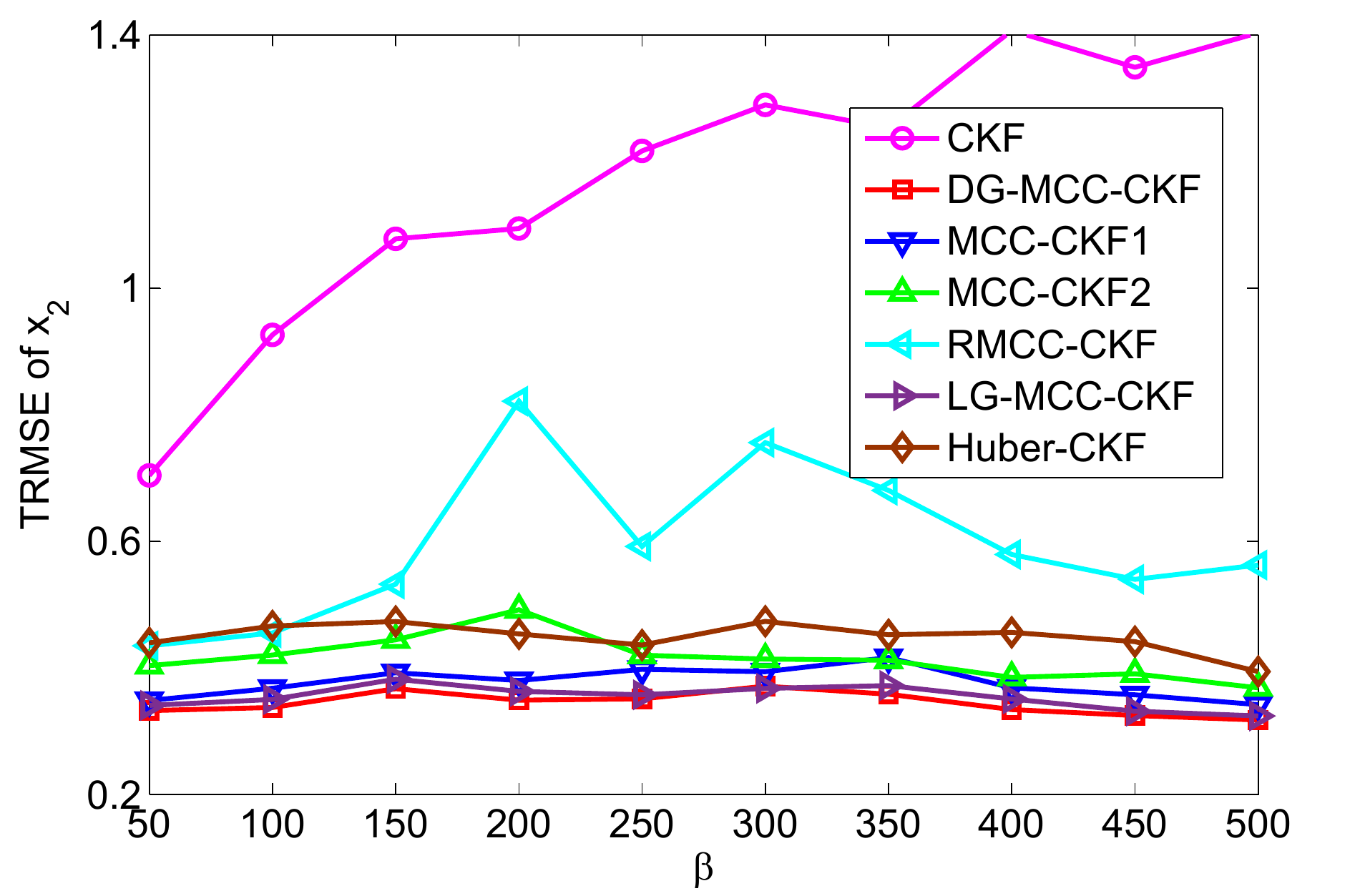}}
\caption{TRMSEs of different algorithms with varying $\varphi$ with fixed $\phi=0.2$.}
\label{f1}
\end{figure*}

\begin{figure*}[th]
\centering
\subfloat[TRMSE of $x_1$]{\includegraphics[width=0.49\columnwidth]{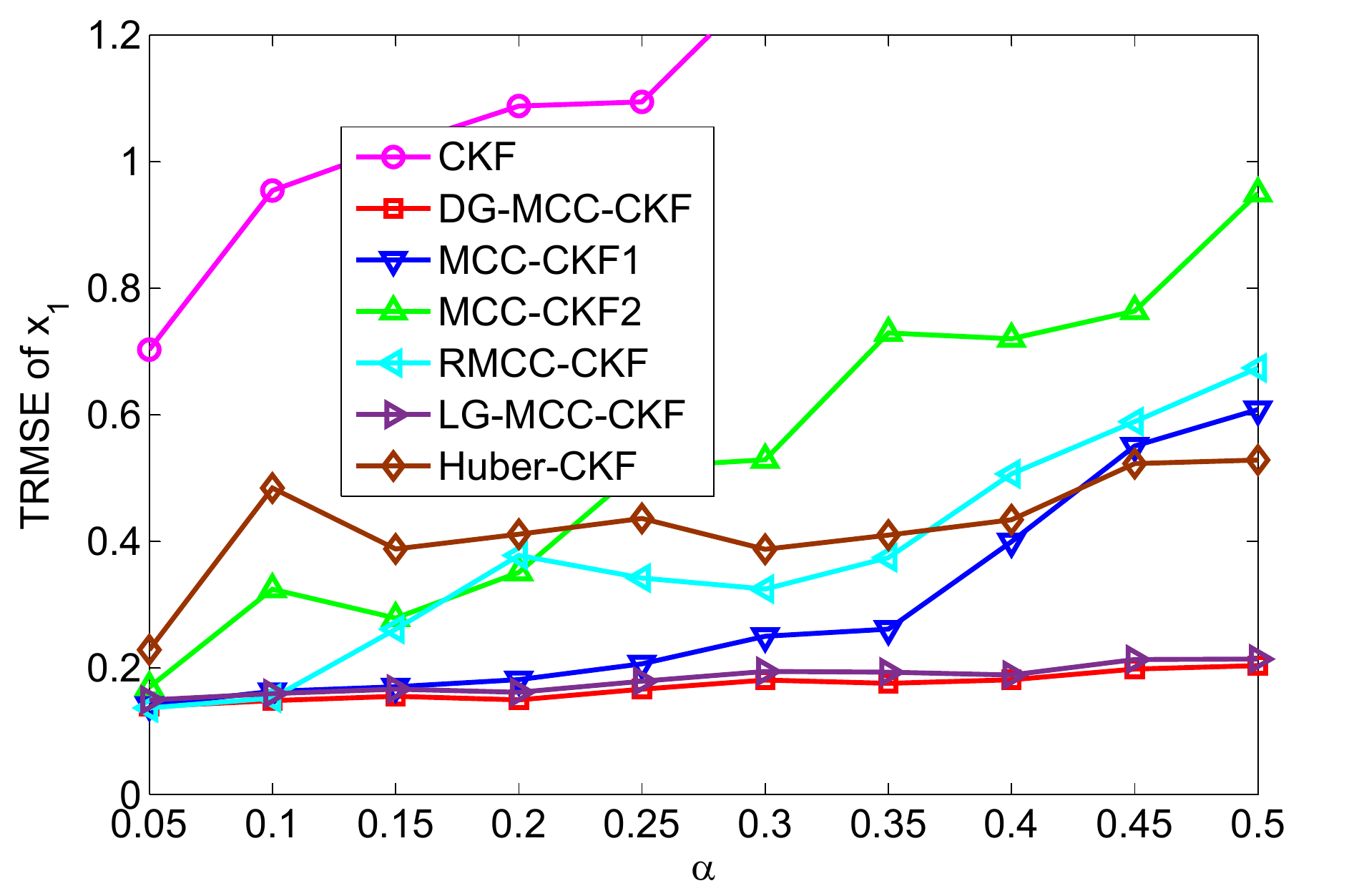}}
\subfloat[TRMSE of $x_2$]{\includegraphics[width=0.49\columnwidth]{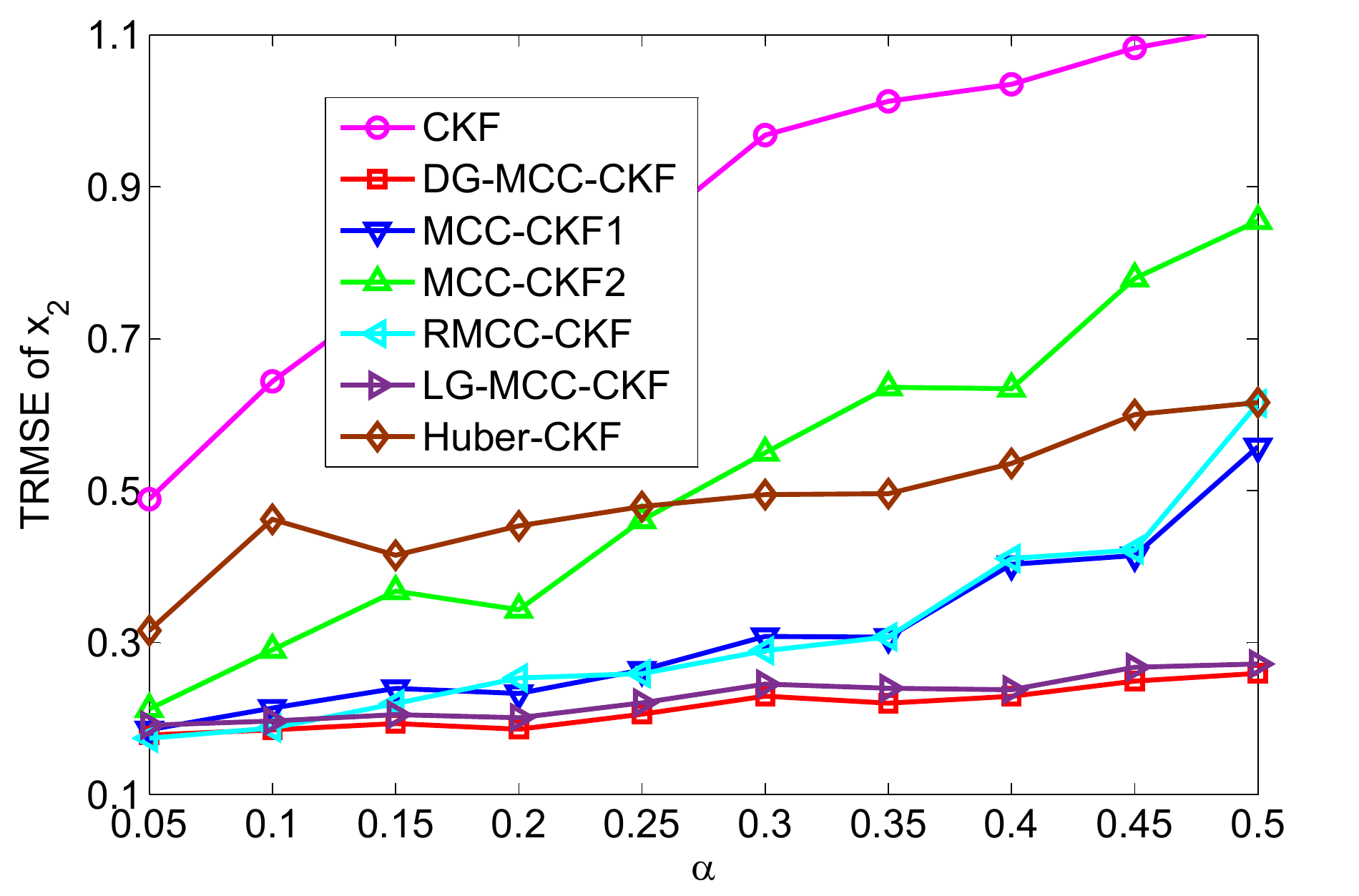}}
\caption{TRMSEs of different algorithms with varying $\phi$ with fixed $\varphi=200$.}
\label{f2}
\end{figure*}

\begin{figure*}[th]
\centering
\subfloat[]{\includegraphics[width=0.49\columnwidth]{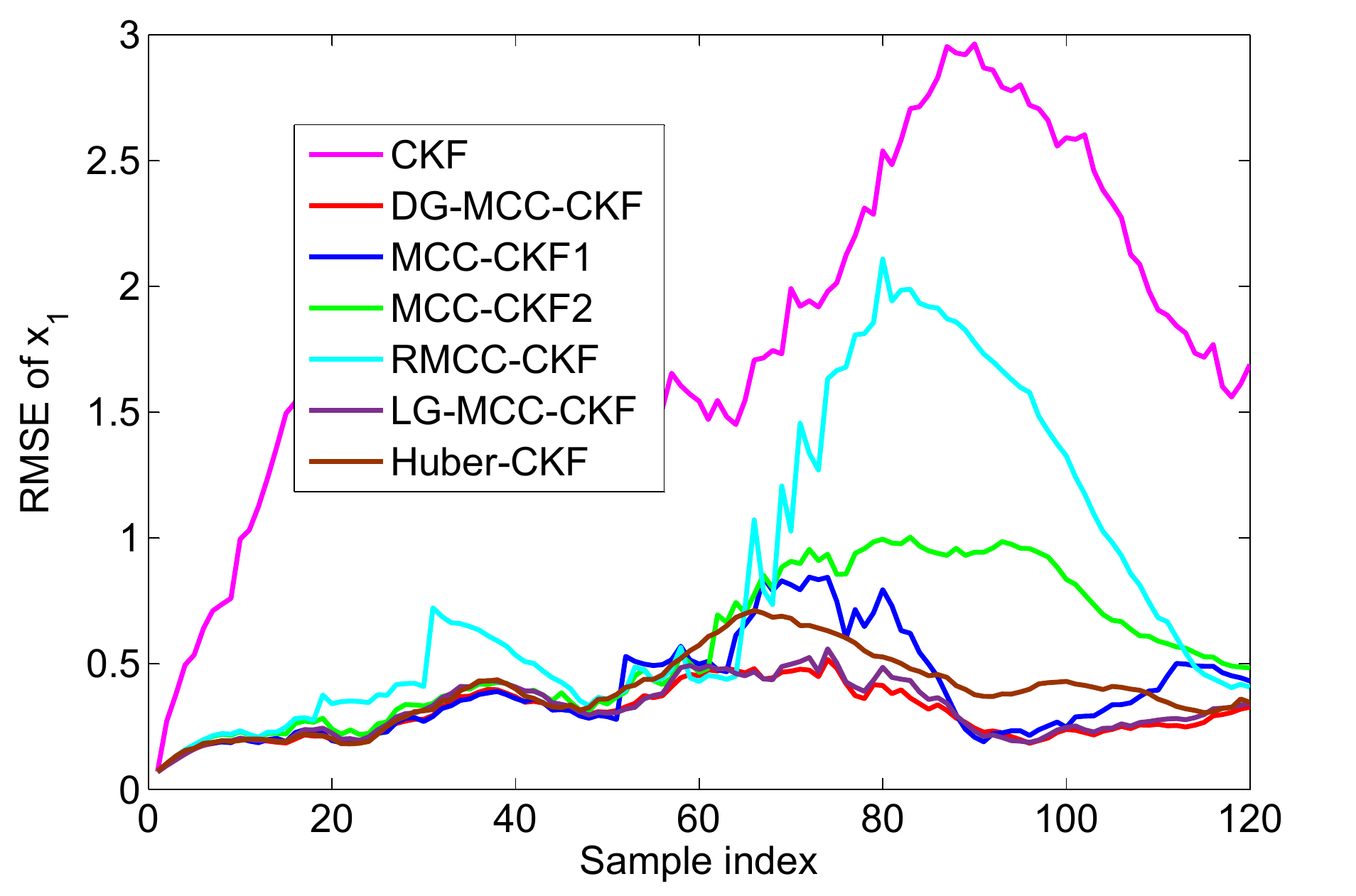}}
\subfloat[]{\includegraphics[width=0.48\columnwidth]{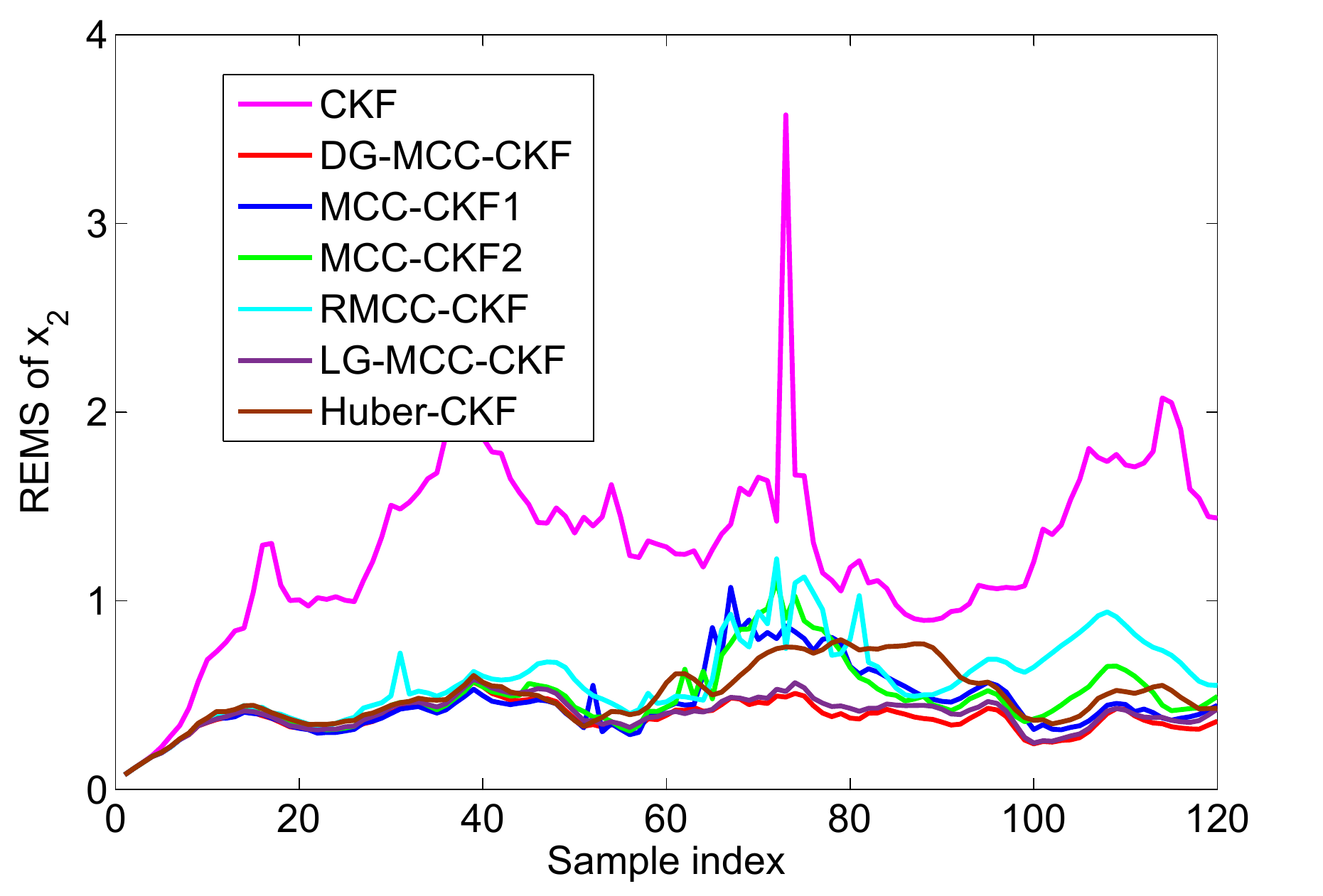}}
\caption{RMSE of $x_1$ and $x_2$ when $\phi=0.3$ and $\varphi=200$.}
\label{f3}
\end{figure*}

Fig.~\ref{f1} illustrates the TRMSEs of $x_1$ and $x_2$ with varying $\varphi$ and fixed $\phi=0.2$; Fig.~\ref{f2} shows these data when $\phi$ varies and $\varphi=200$. It can be seen that, as expected, the conventional CKF degrades significantly since the quadratic loss in the CKF is sensitive to outliers. Overall, our proposed DG-MCL-CKF and LG-MCL-CKF, which have similar performance, have the smallest TRMSEs among all robust solutions, and the RMCC-CKF has the largest ones. The inferior performance of the RMCC-CKF is due primarily to the linearization error during the linear regression procedure. The Huber-CKF performs comparably against to the MCC-CKF, the performance of which is significantly influenced by the kernel parameters. This is illustrated by the fact that the MCC-CKF1 outperforms the MCC-CKF2. Similar conclusions can also be drawn from Fig.~\ref{f3} in which we present the RMSE of the two components of the state for the different algorithms in the scenario where $\phi=0.3$ and $\varphi=200$.

We have further studied how the parameter $\alpha$ influences the performance of the proposed method. Table~\ref{t1} presents the TRMSEs of $x_1$ and $x_2$ in the two selected scenarios with different $\alpha$. We only show the data of the DG-MCL-CKF and omit that of the LG-MCL-CKF due to the similarity. Obviously, the DG-MCL-CKF degrades to the MCC-CKF1 when $\alpha=1$ while it turns to be the MCC-CKF2 when $\alpha=0$. From the results one can observe that the DG-MCL-CKF (i.e., $\alpha\ne 1 \ \text{or}\ \alpha\ne0)$ outperforms both the MCC-CKF1 and MCC-CKF2, so it is concluded that the mixture correntropy is superior over the conventional correntropy. This may bring us a heuristic idea for designing a correntropy related robust Kalman filtering algorithm, i.e., using the mixture correntropy with a larger kernel parameter and a relative small one to alternate the original correntropy to skip the kernel parameter selection step. The optimal value of $\alpha$, however, still needs further investigation.

\begin{table}[!h]
\centering
\caption{TRMSE of $x_1$ and $x_2$ with different $\alpha$}
\label{t1}
\begin{tabular}{ccccccccc}
\hline
$\alpha$       &    & 0(MCC-CKF1) & 0.1    & 0.3    & 0.5    & 0.7    & 0.9    & 1(MCC-CKF2) \\ \hline
\multirow{2}{*}{$\phi=0.3,\varphi=200$ }& $x_1$ & 0.4245      & 0.3692 & 0.3625 & 0.3631 & 0.3650 & 0.3578 & 0.4912      \\ \cline{2-9}
         & $x_2$ & 0.4792      & 0.4243 & 0.4180 & 0.4154 & 0.4154 & 0.4127 & 0.5004      \\ \hline
\multirow{2}{*}{$\phi=0.2,\varphi=300$} & $x_1$ & 0.2726      & 0.2665 & 0.2599 & 0.2544 & 0.2498 & 0.2458 & 0.3640      \\ \cline{2-9}
         & $x_2$ & 0.3522      & 0.3387 & 0.3327 & 0.3280 & 0.3243 & 0.3213 & 0.3709     \\ \hline
\end{tabular}
\end{table}

\subsection{SoC estimation in batteries }

Owing to its high power density, low cost and long cycle life, the lithium-ion battery is widely employed in numerous applications such as electric vehicles. SoC, the level of the amount of charge remaining in a battery, is a crucial monitored parameter in these applications. Unfortunately, SoC is not in general physically measurable. A considerable amount of effort has been devoted to providing an accurate estimate of SoC. One common solution is based on the Kalman filter, in which the evolution of SoC over time is modeled by a nonlinear SSM according to the equivalent circuit of a battery.
\begin{figure*}[th]
\centering
{\includegraphics[width=0.47\columnwidth]{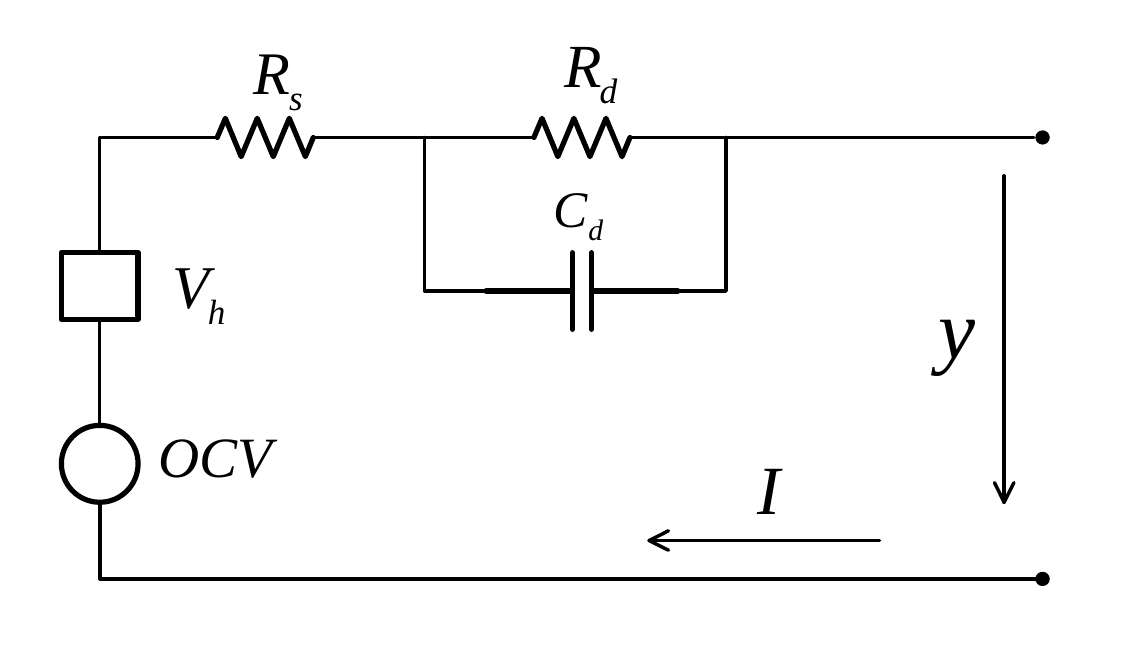}}
\caption{The equivalent circuit of a lithium-ion battery. $V_h$ is a hysteresis voltage source and $OCV$ is the open circuit voltage source.}
\label{f555}
\end{figure*}

Here we consider a equivalent circuit of the lithium-ion battery~\cite{wang2017revisiting}, which is showed in Fig.~\ref{f555}. The associated nonlinear system is given by
\begin{align}
\left\{
\begin{array}{l}
\dot{a} =-\beta I \\
\dot{b} =-\frac{1}{R_{d} C_{d}} b+\frac{I}{C_{d}} \\
\dot{c} =-\gamma I\left[0.0755(1-a)+c\right] \\
y =h(a)-b+c-R_{s} I \\
h(a) = -1.031e^{-35a}+3.685+0.2156a-0.1178a^2+0.3201a^3\\
\end{array}
\right.
\label{bat_con}
\end{align}
where $a$ is SoC; $b$ is the voltage of the RC circuit; $c$ is the hysteresis voltage; $I$ is the discharging current; $y$ is the measurement of the terminal voltage; $\beta$, $R_d$, $C_d$, and $R_s$ are some parameters of the lithium-ion battery. Clearly, the measurement is complicatedly related to SoC, hence outlier-contaminated measurements may influence the estimate accuracy of SoC. We here apply the proposed robust filters to reduce the negative effect of outliers.

In the simulation, denote $\bm x=[a,b,c]^T$ and discretize~\eqref{bat_con} by the Euler method to construct the SSM so that the KF can be applied to estimate SoC. We set $\beta=5.634\time 10^{-5}$, $R_d=3\time 10^{-3}$ ${\Omega}$, $C_d=9\time 10^3$ F, $R_s=5\time 10^{-3}$ ${\Omega}$, $\gamma = 2.47\time 10^{-3}$. The process noise obeys $\mathcal{N}(0,10^{-6}\bm I_3)$, and the measurement noise is from the following Gaussian mixture noise
\begin{align*}
(1-\lambda)\mathcal{N}(0,\bm R)+\lambda\mathcal{N}(0,\kappa\bm R)
\end{align*}
where $\bm R = 10^{-2}$. The true value of the initial state is $\bm x_0=[1,0,0]^T$. all filters are initialized by $\mathcal{N}(\hat{\bm x}_{0|0}, \bm P_{0|0})$ where $\hat{\bm x}_{0|0} = [0.95,0.1,0.001]^T$ and $\bm P_{0|0} = 0.05\bm I_3$.

\begin{figure*}[th]
\centering
\subfloat[Varying $\kappa$ when $\lambda=0.2$]{\includegraphics[width=0.49\columnwidth]{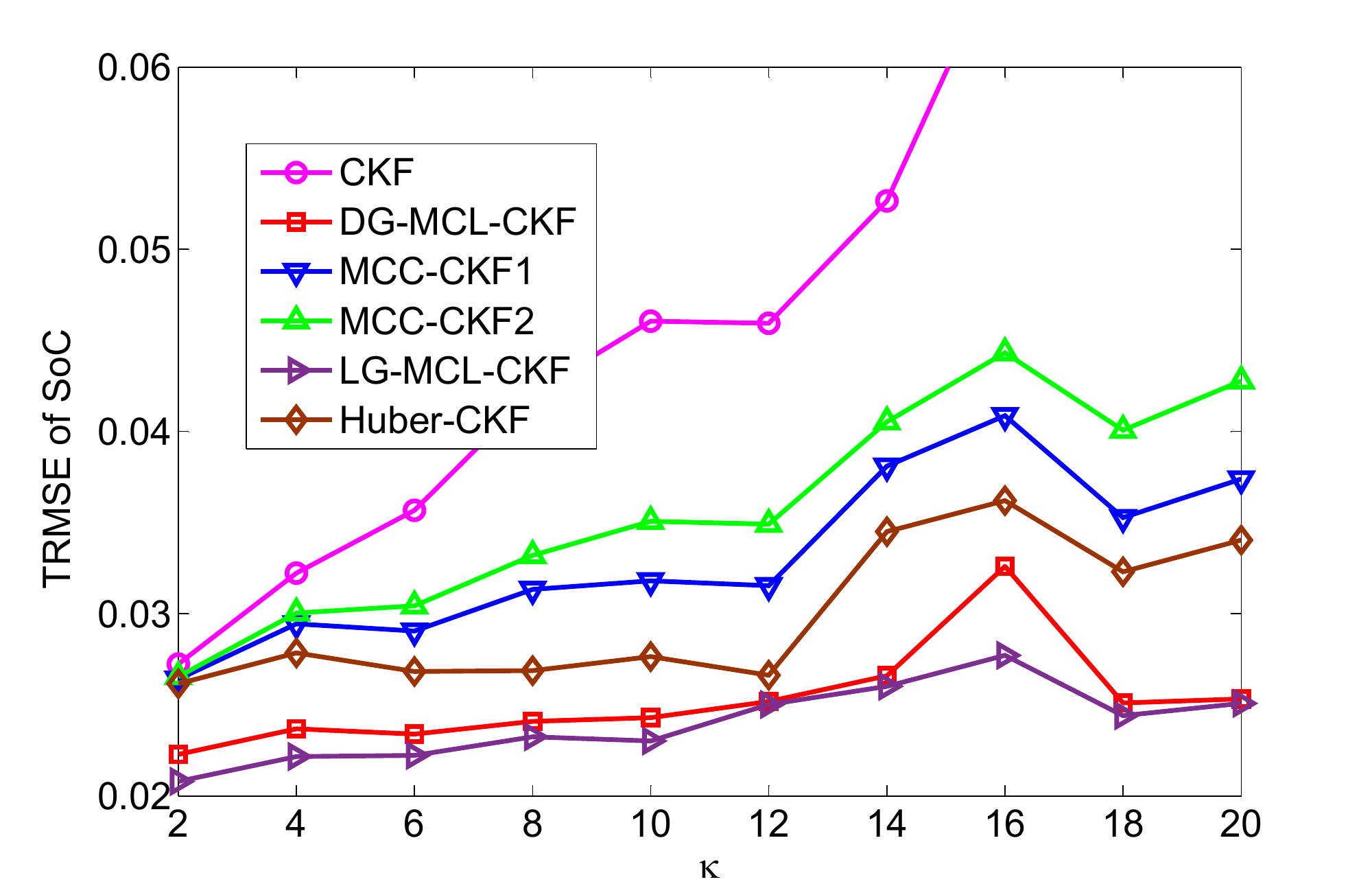}{\label{ff1}}}
\subfloat[Varying $\lambda$ when $\kappa=10$]{\includegraphics[width=0.47\columnwidth]{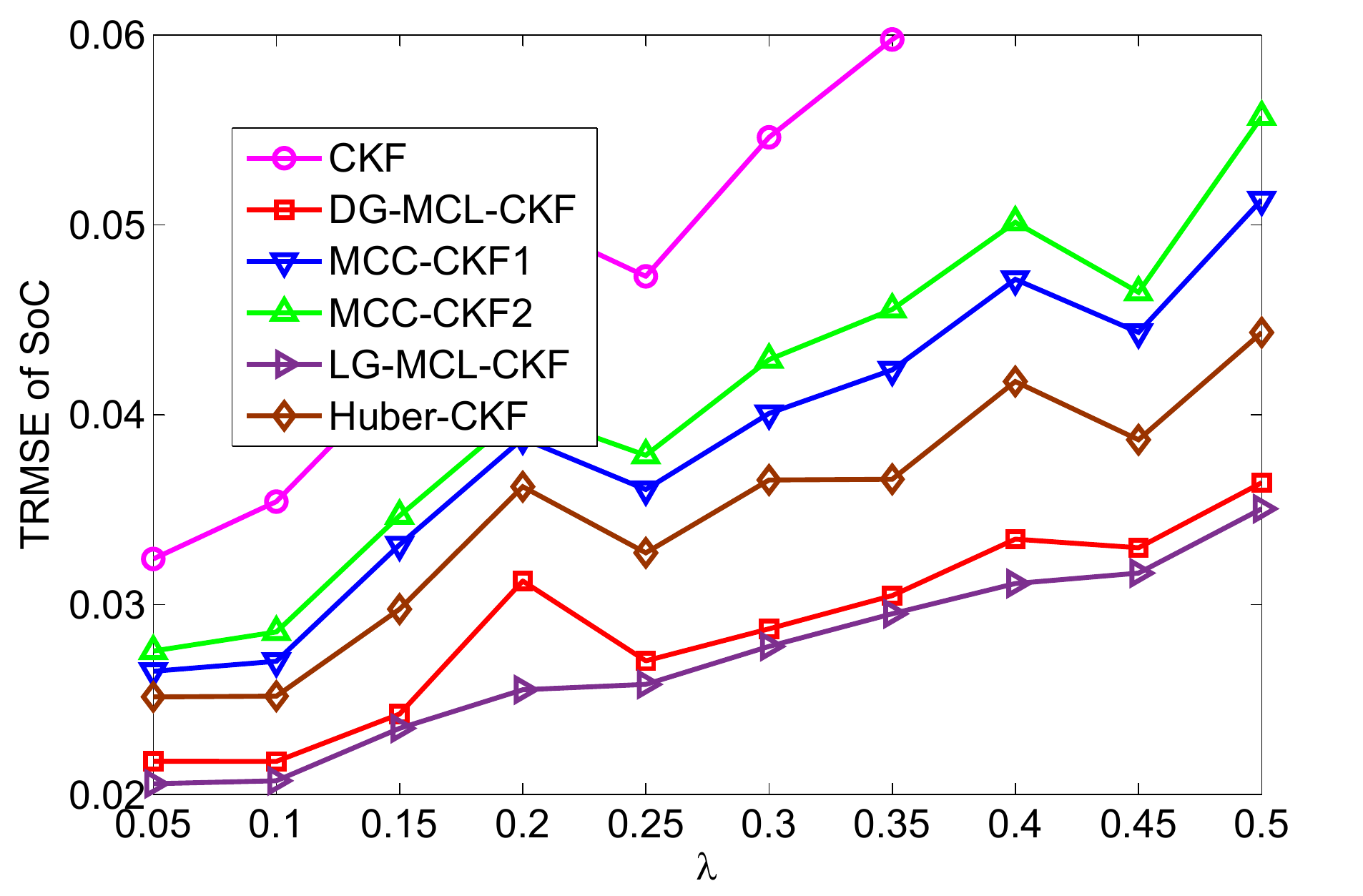}{\label{ff2}}}
\caption{TRMSEs of SoC in two different scenarios.}
\label{f4}
\end{figure*}

\begin{figure*}[th]
\centering
{\includegraphics[width=0.8\columnwidth]{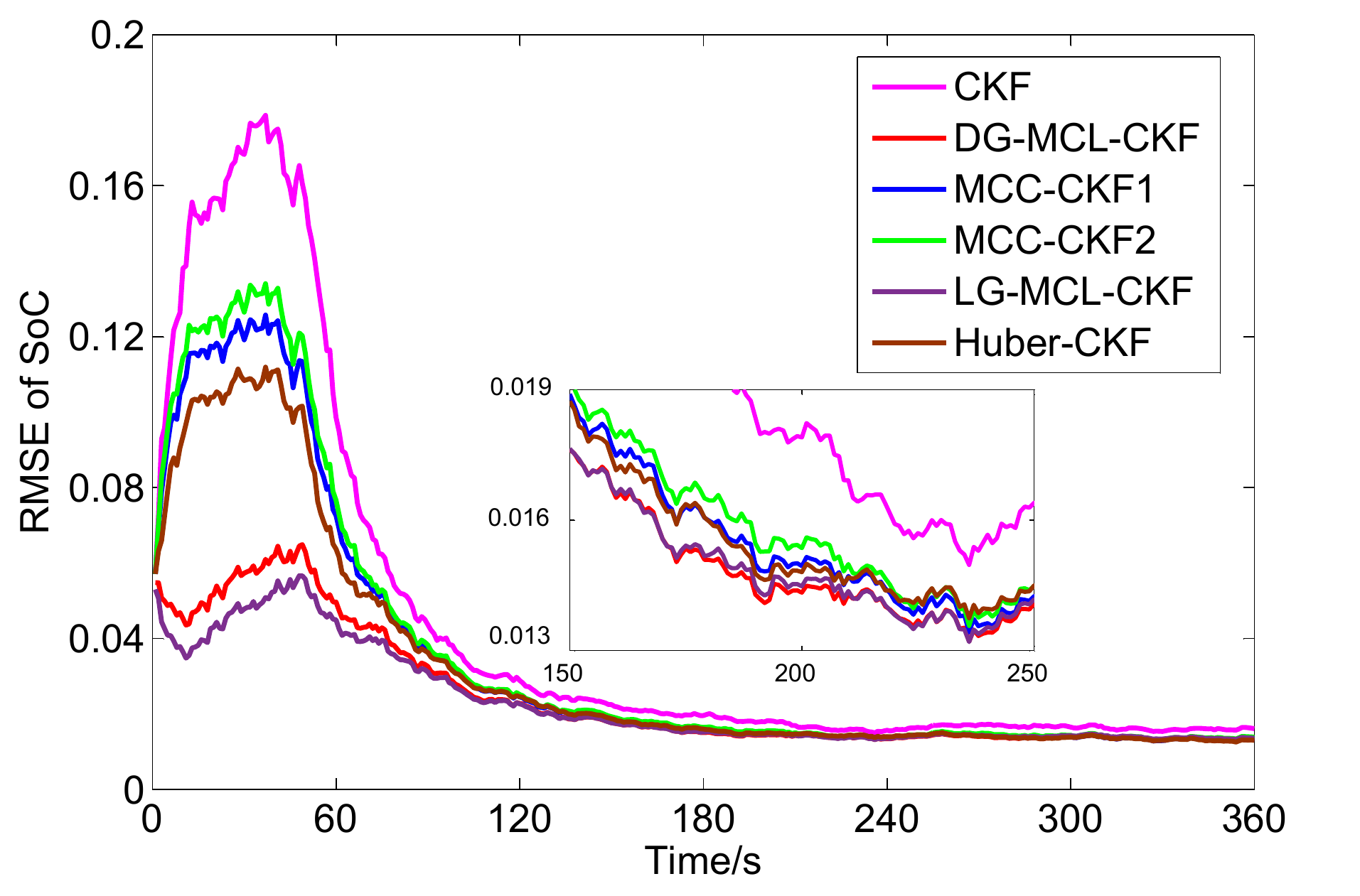}}
\caption{RMSEs of SoC when $\kappa=10$ and $\lambda=0.2$}
\label{f5}
\end{figure*}

The TRMSE and RMSE of SoC, which are based on 100 independent Monte Carlo runs, are utilized as metrics to illustrate the performance of the different filters. The results for different filters are presented in Fig.~\ref{f4} and Fig.~\ref{f5}. The TRMSEs of SoC when $\lambda=0.2$ and $\kappa$ varies are shown in Fig.~\ref{ff1}. Among the robust filters, the proposed MCL based solutions, which perform similarly, have the lowest TRMSE for all $\kappa$. It also can be verified that the TRMSEs of all filters increase slightly when $\kappa$ is small, while fluctuate dramatically for these larger $\kappa$. Fig.~\ref{ff2} shows the SoC TRMSEs versus the change of $\lambda$. It is seen that the performance of all robust filters degrade with the increase of $\lambda$. Again, under such scenarios, our methods outperform other robust filters.

The RMSEs of SoC for the compared robust solutions under a certain scenario are presented in Fig.~\ref{f5}.
Although all filters have converged over time, the convergence speed of our methods is faster than others. It is noted that the convergence values of all robust filters are similar, which are about 30\% smaller than that of the conventional CKF.

\section{Conclusion}
\label{s5}

In this paper, we have investigated outlier-robust Kalman filters based on mixture correntropy for a nonlinear system involving the heavy-tailed measurement noise. Two mixture correntropy induced losses are employed to replace the quadratic loss for the measurement fitting error in the conventional Kalman filtering framework. The resulting robust Kalman filtering problems are then iteratively solved by the conventional CKF with a reweighted covariance matrix of the measurement noise. It can be noted from the simulation results that the proposed algorithms can outperform the existing MCC based solutions.

In the current work, we only consider two kind of Mercer's kernels, i.e., the Gaussian kernel and Laplace kernel. We do not take other kernels such as the Student's t kernel into account, which can be conducted in the further work. In addition, the mixture correntropy based on the multi-kernel method are expected to be the other research direction.

\appendix
\section{Cubature Kalman Filter~\cite{arasaratnam2009cubature}}
\label{appe}

For the state-space model described in~\eqref{process} and~\eqref{measurement} with the Gaussian process and measurement noises, the CKF is implemented as follows:
\begin{enumerate}[Step 1:]
\item Initialize the initial state $\bm x_0\sim\mathcal{N}(\hat{\bm x}_{0|0},\bm P_{0|0})$ and generate the basic weighted cubature point set $\{\bm \xi_{i},\eta_i\}$ for $i=1,\cdots,2n$, where $n$ is the dimension of the state, $\bm \xi_{i}=\sqrt{n}[\bm I]_i$, $[\bm I]=[\bm I_{n},-\bm I_{n}]$ and $\eta_i=1/(2n)$.
\item Generate the sigma points related to the distribution $\mathcal{N}(\hat{\bm x}_{t-1|t-1},\bm P_{t-1|t-1})$
    \begin{align}
    \bm P_{t-1|t-1}&=\bm S_{t-1|t-1}\bm S_{t-1|t-1}^T\\
    \bm\xi_{i,t-1}&=\bm S_{t-1|t-1}\bm\xi_i+\hat{\bm x}_{t-1|t-1}
    \end{align}
\item Calculate the predicted state and its associated error covariance
    \begin{align}
    \bm \chi_{i,t-1}=&f(\bm \xi_{i,t-1})\\
    \hat{\bm x}_{t|t-1}=&\sum_{i=1}^{2n}\eta_i\bm  \chi_{i,t-1}\label{pu_1}\\
    \bm P_{t|t-1}=&\sum_{i=1}^{2n}\eta_i(\bm \chi_{i,t-1}-\bm \hat{\bm x}_{t|t-1})(\bm \chi_{i,t-1}-\bm \hat{\bm x}_{t|t-1})^T+\bm Q_{t-1}\label{pu_2}
    \end{align}
\item Generate the sigma points for the predicted distribution $\mathcal{N}(\hat{\bm x}_{t|t-1},\bm P_{t|t-1})$
    \begin{align}
    \bm P_{t|t-1}&=\bm S_{t|t-1}\bm S_{t|t-1}^T\\
    \bm \phi_{i,t}&=\bm S_{t|t-1}\bm\xi_i+\bm \hat{\bm x}_{t|t-1}
    \end{align}
\item Calculate the predicted measurement, predicted measurement covariance and state-measurement covariance
    \begin{align}
    \bm \psi_{i,t}&=h(\bm \phi_{i,t}),\ \bm{\hat y}_t=\sum_{i=1}^{2n}\eta_i \bm \psi_{i,t} \\
    \bm P_{yy}&=\sum_{i=1}^{2n}\eta_i\left(\bm \psi_{i,t}-\bm{\hat y}_t\right)\left(\bm \psi_{i,t}-\bm{\hat y}_t\right)^T+\bm R_t\label{I_upt} \\
    \bm P_{xy}&=\sum_{i=1}^{2n}\eta_i\left(\bm \chi_{i,t}-\bm{\hat x}_{t|t-1}\right)\left(\bm \psi_{i,t}-\bm{\hat y}_t\right)^T
    \end{align}
\item Obtain the filtered state and its associated error covariance
    \begin{align}
    \hat{\bm x}_{t|t}&=\hat{\bm x}_{t|t-1}+\bm K_t(\bm y_t-\hat{\bm y}_t)\label{x_upt}\\
    \bm P_{t|t}&=\bm P_{t|t-1}-\bm K_t\bm P_{yy}\bm K_t^T\label{p_upt}\\
    \bm K_t&=\bm P_{xy}\bm P_{yy}^{-1}\label{G_upt}
    \end{align}
\end{enumerate}

%\bibliography{mixture_correntropy}

\end{document}